\documentclass[12pt]{article}
\usepackage{amsfonts,amsmath,amssymb,epsf}
\usepackage{graphicx,color}

\newcommand{\f}{\frac}

\newcommand{\om}{\omega}
\newcommand{\al}{\alpha}
\newcommand{\ep}{\epsilon}

\newcommand{\NS}{\mbox{NS}}
\newcommand{\tNS}{\widetilde{\mbox{NS}}}
\newcommand{\R}{\mbox{R}}
\newcommand{\tR}{\widetilde{\mbox{R}}}
\newcommand{\sNS}{\msc{NS}}
\newcommand{\stNS}{\widetilde{\msc{NS}}}
\newcommand{\sR}{\msc{R}}
\newcommand{\stR}{\widetilde{\msc{R}}}

\newcommand{\deebar}{\bar{\partial}}

\newcommand{\lb}{\lbrack}
\newcommand{\rb}{\rbrack}

\newcommand{\msc}[1]{\mbox{\scriptsize #1}}
\newcommand{\dsp}{\displaystyle}

\newcommand{\sgn}{\mbox{sgn}}

\newcommand{\br}{\mbox{{\bf R}}}
\newcommand{\bz}{\mbox{{\bf Z}}}

\newcommand{\bsz}{\msc{{\bf Z}}}

\newcommand{\cN}{{\cal N}}

\newcommand{\cC}{{\cal C}}
\newcommand{\cQ}{{\cal Q}}

\newcommand{\tJ}{\tilde{J}}

\newcommand{\ket}[1]{{|#1\rangle}}
\newcommand{\bra}[1]{{\langle#1|}}
\newcommand{\dket}[1]{{\left.\left|#1\right\rangle\right\rangle}}

\newcommand{\Th}[2]{\Theta_{#1,#2}}
\renewcommand{\th}{{\theta}}

\newcommand{\ch}[2]{\mbox{ch}^{#1}_{#2}}

\newcommand{\nn}{\nonumber\\}

\newcommand {\eqn}[1]{(\ref{#1})}

\makeatletter
\@addtoreset{equation}{section}
\def\theequation{\thesection.\arabic{equation}}
\makeatother
\begin{document}


\begin{titlepage}
 \
 \renewcommand{\thefootnote}{\fnsymbol{footnote}}
 \font\csc=cmcsc10 scaled\magstep1
 {\baselineskip=14pt
 \rightline{
 \vbox{\hbox{hep-th/0406173}
       \hbox{UT-04-18}
       }}}

 \baselineskip=20pt
\vskip 2cm
 
\begin{center}

{\huge Boundary States for the Rolling D-branes}

\vskip 5mm

{\huge in NS5 Background}

 \vskip 2cm
\noindent{ \large Yu Nakayama, Yuji Sugawara and Hiromitsu 
  Takayanagi} \\
{\sf nakayama@hep-th.phys.s.u-tokyo.ac.jp~,~
sugawara@hep-th.phys.s.u-tokyo.ac.jp~,~\\
hiro@hep-th.phys.s.u-tokyo.ac.jp}
\bigskip

 \vskip .6 truecm
 {\baselineskip=15pt
 {\it Department of Physics,  Faculty of Science, \\
  University of Tokyo \\
  Hongo 7-3-1, Bunkyo-ku, Tokyo 113-0033, Japan}
 }
\end{center}

\bigskip

\begin{abstract}

In this paper we construct the time dependent boundary states 
describing the ``rolling D-brane solutions'' in the NS5 background
discovered recently by Kutasov by means of the classical DBI analysis.
We first survey some aspects of non-compact branes 
in the NS5 background based on known boundary states 
in the $\cN=2$ Liouville theory.  
We consider two types of non-compact branes, one of which is BPS and 
the other is non-BPS but stable. Then we clarify how to Wick-rotate 
the non-BPS one appropriately. 
We show that the Wick-rotated boundary state realizes 
the correct trajectory of rolling D-brane in the classical limit, 
and leads to well behaved  spectral densities of open strings 
due to the existence of non-trivial damping factors of energy. 
We further study the cylinder amplitudes 
and the emission rates of massive closed string modes.

\end{abstract}

\vfill

\setcounter{footnote}{0}
\renewcommand{\thefootnote}{\arabic{footnote}}
\end{titlepage}
\baselineskip 18pt


\section{Introduction}

~

Time-dependent physics in string theory is challenging 
and includes many puzzling issues, while important 
for many applications in cosmological problems, for example.
Typical time-dependent processes in string theory 
could be accompanied by radiations of massive as well as 
massless string modes, often giving rise to Hagedorn-like divergences. 
The most naive approach to these problems is to use the low energy effective 
field theories, neglecting the effects of radiations at the stringy level.  
In case of the brane dynamics it is reduced to the supergravity 
coupled to the Dirac-Born-Infeld (DBI) action. 
A direct worldsheet approach to the time-dependent string theory 
is usually much more difficult.  One of the well-studied subjects
is the time evolution of unstable D-branes in the flat 
(or almost trivial) backgrounds, that is, the problem of 
rolling tachyons or s-branes {\em e.g.} \cite{Sen-RT,GS1,Strominger}.

A more ambitious problem may be the time-dependent string 
or brane dynamics in {\em curved} backgrounds. 
A significant background
that is solvable by the exact worldsheet CFT 
is the system of coincident NS5-branes. 
It is familiar that the near horizon physics of NS5 can be well captured by 
the Callan-Harvey-Strominger (CHS) superconformal system \cite{CHS}.
It is expressed as the $SU(2)$ (super) WZW model 
tensored by the linear dilaton theory describing the throat of NS5 \cite{OV}, 
which may be rearranged as the ${\cal N} = 2$ super Liouville theory 
coupled to the ${\cal N} = 2$ minimal model and has been proposed
to be holographically dual to the 6-dimensional Little String Theory
(LST) \cite{ABKS,GKP,GK}.
In this work we potentially make use of known results in 
$(\cN =2)$ Liouville theories and related topics without giving
detailed explanations. See, for instance, \cite{Nakayama} for a review 
and a detailed list of literature.

Recently, Kutasov has studied the time evolution of  
D-branes near the stack of NS5-branes by using the DBI action 
\cite{Kutasov}, emphasizing the formal resemblance to the rolling
tachyon problem. 
It is quite interesting that the radial position 
of D-brane plays the similar roles to the rolling tachyon field. 
He found a time-dependent solution which we shall call the ``rolling
D-brane" in this paper. That looks like
\begin{eqnarray}
 e^{-\f{\cQ\phi}{2}}=\f{\tau_p}{E}\cosh \f{\cQ t}{2}~,
\label{rolling solution}
\end{eqnarray}
where $\cQ = \sqrt{2/N}$ is the amount of the linear dilaton
($N$ is the NS5 charge) and $\tau_p$, $E$ are  
the tension, energy of D$p$-brane. 
After emitted from the NS5-branes in the infinite past, 
the rolling D-brane is attracted by the gravitational force from NS5,  
and reaches the maximum position in the radial direction at a certain time.  
Then it is eventually reabsorbed into the NS5-branes.
He has pointed out that, after making the Wick rotation, 
the orbit of rolling brane is identified as the ``hairpin brane" solution 
presented in \cite{LVZ} \footnote{
The closely related $D1$-brane solution in the $SL(2;\br)/U(1)$ coset CFT 
was constructed in \cite{RibS} earliar than \cite{LVZ} was published. }
,
which is constructed in a bosonic CFT of two scalar fields.

In this paper, inspired by these observations, we study 
aspects of the rolling D-branes by the methods of boundary 
conformal field theory (BCFT). 
In order to construct the boundary states for the rolling branes,
we start from the known boundary states describing 
non-compact branes in the ${\cal N} = 2$ Liouville theory.
It is simply regarded as the supersymmetrized version 
of the hairpin brane of \cite{LVZ}. 
Indeed, the authors of \cite{LVZ} have introduced the screening charges, 
which are just regarded as the sine-Liouville type potentials. 
Then they illustrated that the chiral algebra compatible 
with these screening charges is the W-algebra, of which  
their hairpin solution has been made up. 
The supersymmetrized version of sine-Liouville theory 
is known to be the $\cN=2$ Liouville theory,  
and the W-algebra used in \cite{LVZ} 
should be replaced with the $\cN=2$ superconformal algebra (SCA). 
Therefore, observing the manifest resemblance of their boundary wave 
functions, it is quite natural to expect that the hairpin branes 
in the ${\cal N} =2$ Liouville theory are identified as 
the ``class 2 branes'' introduced in \cite{ES-L,ASY}.\footnote
  {See also \cite{MMV,IPT2,ASY2,FNP} for closely related studies.}
They are defined associated to the non-degenerate representations
and regarded as the ${\cal N} = 2$ extension of 
the FZZT branes \cite{FZZT}. We will directly clarify this point 
by analysing the position space boundary wave functions. 
Even though the NS sector of our hairpin brane has the almost same 
structure as the bosonic one \cite{LVZ}, the analysis on the R sector
sheds new light on the physics of non-BPS D-branes in the NS5
backgrounds.

We then try to make the (inverse) Wick rotation 
of the hairpin branes into time-dependent boundary states. 
As we will see, the naive Wick rotation  in the momentum space 
does not work. 
Actually, if we do it, due to the bad UV behavior
we can obtain neither a sensible 
open channel spectrum nor the correct classical orbit of rolling brane. 
We propose  how one should correctly perform the Wick rotation, 
and show that 
\begin{description}
 \item[(1)] we obtain the expected trajectory of 
 rolling D-brane solution in the classical limit, 
 \item[(2)] we obtain well-defined spectral densities of open strings 
as opposed to the naive rotation.
\end{description}
Interestingly, our procedure of Wick rotation 
gives rise to a damping factor of energy 
similar to the prefactor characteristic for the rolling tachyon solution
\cite{Sen-RT}. 
It yields an improved UV behavior that makes it possible 
to gain  the well-defined  spectral densities. 
We further analyse the cylinder amplitudes and 
the rates of the closed string emissions in the rolling process.

~

Throughout this paper we shall use the convention $\al'=2$
and set $q\equiv e^{2\pi i \tau}$ as usual. 
$T (\in \br_{>0})$ and $t (\equiv 1/T)$ will be used as 
the closed and open channel moduli respectively
in cylinder (annulus) amplitudes. 

~

\section{Non-compact Static Branes in NS5 Backgrounds}

~

In this section we study two types of static D-branes in 
the NS5 background, one of which is BPS and the other is non-BPS but
stable. Throughout this section we choose the Neumann boundary 
condition along the time direction.

It is familiar that the stack of $N$ NS5-branes is 
described by the CHS superconformal system 
\cite{CHS} in the near horizon limit;
\begin{eqnarray}
\br^{5,1} \times \br_{\phi} \times SU(2)_{N-2} \cong
\br^{5,1} \times \frac{\left\lb \br_{\phi} \times S^1_Y \right\rb \times
  M_{N-2}}{\bz_{N}}~,
\label{CHS}
\end{eqnarray} 
where $M_{k}$ denotes the $\cN=2$ minimal model with level $k$
($\hat{c}=k/(k+2)$) and $\br_{\phi} \times S^1_Y$ means the 
$\cN=2$ Liouville theory with $\hat{c}=1+\cQ^2 = 
1+\frac{2}{N}$. The $\bz_N$-orbifolding is defined 
with respect to the total $\cN=2$ $U(1)$-charge as in 
the Gepner models and assures the space-time SUSY. 
The criticality condition is satisfied as 
\begin{eqnarray}
\left(1-\frac{2}{N}\right) + \left(1+\frac{2}{N}\right) = 2~,
\end{eqnarray}
along the transverse direction of NS5. 
The compact boson $Y$ has radius $\cQ$ and roughly identified with 
the $J^3$-direction of $SU(2)$-WZW model. 
The linear dilaton is parametrized as $\Phi(\phi)=
-\frac{\cQ}{2}\phi$ in our convention. 
Namely, $\phi\,\sim\, +\infty$
is the weakly coupled asymptotic region, and $\phi\,\sim\, -\infty$
is the strongly coupled region near the NS5-branes.
The $\cN=2$ superconformal currents are explicitly written as
\begin{equation}
\left\{
\begin{array}{l}
\dsp  T =-\frac{1}{2}(\partial Y)^2 
-\frac{1}{2}(\partial \phi)^2 -\frac{\cQ}{2}\partial^2\phi
-\frac{1}{2}(\Psi^+\partial \Psi^- -\partial \Psi^+ \Psi^-) \\
\dsp  G^{\pm}=-\frac{1}{\sqrt{2}}\Psi^{\pm}(i\partial Y \pm \partial \phi )
\mp \frac{\cQ}{\sqrt{2}}\partial \Psi^{\pm} \\
\dsp J= \Psi^+\Psi^- - \cQ i\partial Y~~,
\end{array}
\right.
\label{SCA-L}
\end{equation}
where $\dsp Y(z)Y(0)\sim -\ln z $, $\dsp \phi(z)\phi(0)\sim -\ln z$, 
$\dsp  \Psi^{\pm}(z)\Psi^{\mp}(0)\sim \frac{1}{z}$, 
$\dsp  \Psi^{\pm}(z)\Psi^{\pm}(0) \sim 0 $,
and $\dsp  \Psi^{\pm}=-\frac{1}{\sqrt{2}}
(\psi^Y\pm i\psi^{\phi})$.

Without any marginal deformations, that is, treating 
as a free conformal theory, 
the stack of NS5-branes leads to a singularity 
at which the dilaton blows up. 
A better treatment is  to incorporate the $\cN=2$ Liouville type 
potential that prevents strings from propagating in the strong coupling 
region.  
The standard choice of Liouville potential is the chiral one;
\begin{eqnarray}
&& \mu \int d^2z\, \Psi^{-}\tilde{\Psi}^{-}
e^{-\frac{1}{\cQ}(\phi+ i Y)} + \mbox{(c.c)}~, 
\label{cosm term 1} 
\end{eqnarray}
which is naturally regarded as the supersymmetrized version 
of sine-Liouville type potential.
This deformation amounts to distributing the NS5-branes 
at equal distances on a circle of radius 
$r\sim |\mu|^{1/N}$ (see {\em e.g.} \cite{GK}).
It breaks the $SU(2)$-rotation symmetry in 
the CHS model, but preserves the $\cN=2$ superconformal symmetry 
\eqn{SCA-L}.

~


\subsection{BPS Non-compact Branes in the NS5 Background}

~

We first study the BPS D-branes with 
non-compact worldvolumes. 
All the things here are compatible 
with the marginal deformation \eqn{cosm term 1}.
We consider the boundary states constructed as 
\begin{itemize}
 \item {\bf $M_{N-2}$-sector :} (A-type) Cardy states 
$\ket{L,M}^{(\sigma)}$,  $\sigma=\NS,\,\R$,~ 
$L=0,\ldots, N-2$, ~ $M\in \bz_{2N}$, 
$L+M \in 2\bz$. (See {\em e.g.} \cite{RS}.)
 \item {\bf $\br_{\phi}\times S^1_Y$-sector : }
The (A-type) ``class 2 states'' constructed in \cite{ES-L,ASY}
which corresponds to the extended massive 
characters ($\cQ= \sqrt{2/N}$);
\begin{eqnarray}
&& \ket{B;P,M'}^{(\sigma)} = \int_0^{\infty}dp\, \sum_{m \in \bsz_{2N}}\,
\Psi^{(\sigma)}_{P,M'}(p,m)\dket{p,m}^{(\sigma)}~, \nn
&& \Psi_{P,M'}^{(\sigma)}(p,m) =
\cQ^{3/2}\, e^{-2\pi i \frac{M'm}{2N}} 
\cos(2\pi P p)\, 
\frac{\Gamma(-i\cQ p)\Gamma\left(1-i\frac{2p}{\cQ}\right)}
{\Gamma\left(\frac{1}{2}+\frac{m-\nu(\sigma)}{2}-i\frac{p}{\cQ}\right)
\Gamma\left(\frac{1}{2}-\frac{m-\nu(\sigma)}{2}-i\frac{p}{\cQ}\right)}~, \nn
&&
\label{class 2 brane}
\end{eqnarray}
where we set $\nu(\sigma)=0,1$ for $\sigma=\NS,\,\R$ respectively,
and omitted the phase factor depending on the (renormalized) cosmological 
constant. The R-sector boundary wave function is defined by the
$1/2$-spectral flow.
$\dket{p,m}^{(\sigma)}$ denotes the A-type Ishibashi states
(namely, Dirichlet along $Y$) associated to the extended massive 
character $\chi^{(\sigma)}(p,m)$ defined by \cite{ES-L}
\begin{eqnarray}
&& \chi^{(\sigma)}(p,m;\tau,z)
\equiv
 q^{\frac{p^2}{2}}\Th{m}{N}
\left(\tau,\frac{2z}{N}\right)\, \frac{\th_{\lb \sigma \rb}
(\tau,z)}{\eta(\tau)^3}~, 
\label{massive character} 
\end{eqnarray}
where $\theta_{\lb \sigma \rb}$ denotes $\th_3$, $\th_4$, $\th_2$, 
$i\th_1$ for $\sigma = \NS, \,\tNS, \, \R, \, \tR$ respectively. 
\end{itemize} 
The desired BPS brane is given by
\begin{eqnarray}
\ket{B;L,M,P,M'}^{(\sigma)}= \cN P_{\msc{closed}} \, \left\lb 
\ket{L,M}^{(\sigma)} \otimes \ket{B; P,M'}^{(\sigma)}
\right\rb~,
\label{BPS brane}
\end{eqnarray}
where $\cN$ is a normalization constant\footnote
  {This normalization constant $\cN$ should be determined 
   by the Cardy condition for the cylinder amplitudes including 
   the open strings that belongs to the {\em discrete} representations.
   We can determine it as $\cN= \sqrt{N}$ by considering the overlaps with  
   class 1 states as in \cite{ES-L}.  In any case its explicit value
   is not important for the analysis in this paper.}
and $P_{\msc{closed}}$ 
is the projection operator to the correct closed string spectrum
in the $\bz_N$-orbifolded theory.  
Especially, $P_{\msc{closed}}$
restricts the total $U(1)$-charge to be an integer. 
Note that, with the existence of $P_{\msc{closed}}$, 
the boundary state $\ket{B;L,M,P,M'}^{(\sigma)}$
actually depends  only on the sum $M+M'$. 
We may thus simply set $M'=0$, and express it as
$\ket{B;L,M,P}^{(\sigma)}$ from here on.

It is instructive to calculate their overlaps (cylinder amplitudes)
explicitly.  
A non-trivial point is the insertion of $P_{\msc{closed}}$, 
which is translated 
into the spectral flow sum in the dual open string channel. 
The next identity is useful in the following calculations;
\begin{eqnarray}
\left|\cosh \pi \left(\frac{p}{\cQ}+ i \frac{m}{2}\right)\right|^2
= \frac{1}{2} \left\{
\cosh \left(\frac{2\pi p}{\cQ}\right) + \cos \left(\pi m\right)
\right\}~.
\label{identity 1}
\end{eqnarray}
Note that, for the $\R$-sector, the shift $m\,\rightarrow\,m-1$
leads to an extra minus sign for  the cosine term. 

The desired overlap is then calculated (up to overall normalization) as 
($t\equiv 1/T$, $\sigma = \NS,\,\R$) 
\begin{eqnarray}
&& {}^{(\sigma)}\bra{B;L_1,M_1,P_1} e^{-\pi T H^{(c)}} 
\ket{B;L_2,M_2,P_2}^{(\sigma)} =
\int_0^{\infty} dp\, \sum_{\ell=0}^{N-2}\, 
\sum_{\stackrel{m\in \bsz_{2N}}{m+\ell\in 2\bsz}}\,
N_{L_1,L_2}^{\ell} \, \ch{(S\cdot \sigma)}{\ell,M_2-M_1-m}(it,0) \nn
&& \hspace{1cm} \times
\left\{
\rho_1(p|P_1,P_2) \chi^{(S\cdot \sigma)}(p,m;it,0)
+ (-1)^{\nu(\sigma)} 2\rho_2(p|P_1,P_2) \chi^{(S\cdot \sigma)}(p,m+N;it,0)
\right\} ~, \nn
&& {}^{(\sigma)}\bra{B;L_1,M_1,P_1} e^{-\pi T H^{(c)}} e^{\frac{i\pi}{2}
(J_0+\tJ_0)}
\ket{B;L_2,M_2,P_2}^{(\sigma)} \nn
&& \hspace{1cm} = (-1)^{\nu(\sigma)}
\int_0^{\infty} dp\, \sum_{\ell=0}^{N-2}\, 
\sum_{\stackrel{m\in \bsz_{2N}}{m+\ell\in 2\bsz 
+ 1}}\,
N_{L_1,L_2}^{\ell} \, \ch{(S\cdot \tilde{\sigma})}{\ell,M_2-M_1-m}(it,0) \nn
&& \hspace{1cm} \times
\left\{
\rho_1(p|P_1,P_2) \chi^{(S\cdot \tilde{\sigma})}(p,m;it,0)
+ (-1)^{\nu(\sigma)}
2\rho_2(p|P_1,P_2) \chi^{(S\cdot \tilde{\sigma})}(p,m+N;it,0)
\right\} ~,
\label{overlap BPS}
\end{eqnarray} 
where we set
\begin{eqnarray}
S\cdot \NS = \NS~,~~ S\cdot \tNS = \R~, ~~ S \cdot \R = \tNS~, ~~ 
S\cdot \tR = \tR~~,
\end{eqnarray}
and $N_{L_1,L_2}^{\ell}$ is the fusion coefficient of $SU(2)_{N-2}$.
(The $\tR$-sector amplitude trivially vanishes.)
The open channel spectral densities $\rho_i(p|P_1,P_2)$ are evaluated as
follows;
\begin{eqnarray}
&& \rho_1(p|P_1,P_2) = \int_0^{\infty}dp'\, \frac{\cos(2\pi p p')}
{\sinh(\pi \cQ p')\sinh\left(\frac{2\pi p'}{\cQ}\right)} \,
\sum_{\ep_i=\pm 1} \, \cosh \left(
2\pi \left(\frac{1}{\cQ}+i\ep_1 P_1 + i\ep_2 P_2\right)p' \right)~, \nn
&&
\label{rho 1} \\
&& \rho_2(p|P_1,P_2) = \int_0^{\infty}dp'\, \frac{\cos(2\pi p p')}
{\sinh(\pi \cQ p')\sinh\left(\frac{2\pi p'}{\cQ}\right)} \,
\sum_{\ep=\pm 1} \, \cos \left(
2\pi \left(P_1 + \ep P_2\right)p' \right)~,
\label{rho 2}
\end{eqnarray}
which can be expressed by the q-Gamma functions \cite{FZZT} 
after subtracting the IR divergences at $p'=0$. 
The remark below \eqn{identity 1} is the origin 
of the extra phase $(-1)^{\nu(\sigma)}$ for the $\rho_2$-terms.

We can show that the brane configuration is supersymmetric, 
if (and only if) $M_1\equiv M_2~(\mod\, 2N)$ is satisfied. 
We demonstrate this fact by observing a simple example: the self-overlap
of $\ket{B;L=0,M,P=0}^{(\sigma)}$. 
The relevant calculation is as follows;
\begin{eqnarray}
&& Z \equiv {}^{(\sNS)}\bra{B;0,M,0}e^{-\pi T H^{(c)}} \ket{B;0,M,0}
^{(\sNS)} \cdot \left(\frac{\th_3}{\eta}\right)^2 (iT) \nn
&& \hspace{2cm}
- {}^{(\sNS)}\bra{B;0,M,0}e^{-\pi T H^{(c)}} 
e^{i\pi \frac{1}{2}(J_0+\tJ_0)}\ket{B;0,M,0}
^{(\sNS)} \cdot \left(\frac{\th_4}{\eta}\right)^2 (iT) \nn
&& \hspace{2cm}
-  {}^{(\sR)}\bra{B;0,M,0}e^{-\pi T H^{(c)}} \ket{B;0,M,0}
^{(\sR)} \cdot \left(\frac{\th_2}{\eta}\right)^2 (iT) \nn
&& = \sum_{\sigma} \ep(\sigma)\, 
\sum_{m\in \bsz_{2N} \cap (2\bsz+\nu(\sigma))}
\, \int_0^{\infty}dp\,\ch{(\sigma)}{0,-m}(it,0)
\left\{ \rho_1(p) \,
\chi^{(\sigma)}(p,m;it,0) \cdot 
\left(\frac{\th_{\lb \sigma \rb}}{\eta}\right)^2 (it)
\right. \nn
&& \hspace{2cm} \left. + (-1)^{\nu(S\cdot \sigma)}2\rho_2(p) \, 
\chi^{(\sigma)}(p,m+N;it,0) \cdot 
\left(\frac{\th_{\lb \sigma \rb}}{\eta}\right)^2 (it)
\right\}~,
\label{self-overlap 1}
\end{eqnarray}
where we set $\ep(\NS) (= \ep(\tR))=1$, $\ep(\tNS)=\ep(\R)=-1$. 
We here implicitly incorporated the contributions from free fermions
in the flat space-time. 
The spectral densities here are written explicitly as 
\begin{eqnarray}
&& \rho_1(p) \equiv \rho_1(p|0,0) = 4\int_0^\infty dp' 
\frac{\cos(2\pi pp')\cosh\left(\frac{2\pi p'}{\cQ}\right)}
{\sinh\left(\pi \cQ p'\right)\sinh\left(\frac{2\pi p'}{\cQ}\right)} ~,
\label{rho 1 0} \\
&& \rho_2(p) \equiv \rho_2(p|0,0)
= 2 \int_0^\infty dp' \frac{\cos(2\pi pp')}{\sinh\left(\pi \cQ p'\right)\sinh\left(\frac{2\pi p'}{\cQ}\right)} \ .
\label{rho 2 0}
\end{eqnarray}
Using \eqn{field identification} (take care of the relative minus sign
in the $\tNS$-character) and the branching relation 
\eqn{branching minimal}, we further obtain
\begin{eqnarray}
&& Z = \int_0^{\infty}dp \, 
\frac{e^{-2\pi t \cdot \frac{p^2}{2}}}{\eta(i t)}\,
\left( \rho_1(p) \, 
\chi^{(N-2)}_0(it)+2\rho_2(p) \chi^{(N-2)}_{N-2}(it)\right) \nn
&& \hspace{2cm} \times
\left\{\left(\frac{\th_3}{\eta}\right)^4(it)
-\left(\frac{\th_4}{\eta}\right)^4(it)
-\left(\frac{\th_2}{\eta}\right)^4(it)
\right\} \equiv 0~,
\label{self-overlap 2}
\end{eqnarray} 
where $\chi^{(N-2)}_{\ell}(\tau)$ is the spin $\ell/2$ 
character of $SU(2)_{N-2}$. 
It is easy to generalize this evaluation to the cases with 
general $L_i$, $P_i$ 
and we again obtain vanishing cylinder amplitudes. 
On the other hand, as discussed in \cite{HNS}, 
in all the cases of $M_1 \not\equiv M_2$ $(\mbox{mod}\, N)$,
the open channel characters include fractional $U(1)$-charges
and do not cancel with one another.
Moreover, in the cases of $M_1 \equiv M_2+N $ $(\mbox{mod}\, 2N)$
we find the inverse GSO projection in the open channel.

~

\subsection{Stable Non-BPS Branes in the NS5 Background : Hairpin Branes}

~

We next study the boundary states that are similar to \eqn{BPS brane} 
but have a significant difference in the physical interpretation.
The idea is very simple: We shall replace the compact boson $Y$,
which is transverse to NS5, with a non-compact (space-like) boson $X$
parallel to NS5. We later try to perform the Wick-rotation of 
$X$ to the time coordinate $X^0$. A subtlety emerges since this formal 
replacement leads to the {\em other\/} $\cN=2$ superconformal currents 
defined with respect to $X$, $\phi$ rather than $Y$, $\phi$, which are 
not compatible with the Liouville potential \eqn{cosm term 1}.
Therefore, we shall consider the model with no marginal deformation 
for the time being. In other words we consider the truly 
`stacked'  NS5-brane configuration.
We later discuss how we should incorporate a marginal deformation 
that avoids singularity and is compatible with 
our D-brane solutions.

In order to achieve the desired boundary states, 
all we have to do is to take the decompactification limit 
in \eqn{class 2 brane}, which makes the $U(1)$-charge continuous.
The $SU(2)$-part is now completely decoupled and we omit it here.
One may tensor any Cardy state in the (super) $SU(2)_{N-2}$-sector  
$\ket{L}$ that is quite familiar \cite{Cardy}. 
For example, $\ket{L=0}$ ($\ket{L=N-2}$)
corresponds to a D0-brane located on the 
north (south) pole on $S^3$ (see {\em e.g.} \cite{alek1,bds}). 
The desired boundary state is written as
\begin{eqnarray}
|B; P,Q\rangle^{(\sigma)} = \int_0^\infty dp\int_{-\infty}^{\infty}d\al \, 
\Psi_{P,Q}^{(\sigma)} (p,\al) \dket{p,\al}^{(\sigma)} \ ,
\label{hairpin brane}
\end{eqnarray}
where the Ishibashi states $\dket{p,\al}^{(\sigma)}$ are defined 
associated to the irreducible massive characters
\begin{eqnarray}
\ch{(\sigma)}{}(p,\al;\tau,z) = q^{\frac{p^2}{2} + \frac{\al^2}{2 \cQ^2}}
 e^{2\pi i \al z} \frac{\th_{\lb \sigma \rb}(\tau,z)}{\eta(\tau)^3} ~.
\end{eqnarray}
It is useful to introduce the rescaled $U(1)$ charge 
as $q = \frac{\al}{\cQ}$ so that the conformal weight becomes 
$ h = \frac{p^2}{2} + \frac{q^2}{2}+ \frac{\cQ^2}{8}$ 
for the NS sector. 
We also change the normalization of the Ishibashi states accordingly.
Then the boundary wave function is  given by
\begin{eqnarray}
&& \Psi_{P,Q}^{(\sigma)}(p,q) = \sqrt{2}\cQ
e^{-2\pi i \frac{Q q}{\cQ}} \cos(2\pi Pp) 
\frac{\Gamma\left(-i\cQ p\right)
\Gamma\left(1-i\frac{2p}{\cQ}\right)}
{\Gamma\left(\frac{1}{2}-i\frac{p}{\cQ}
+\frac{q}{\cQ} -\frac{\nu(\sigma)}{2}\right)
\Gamma\left(\frac{1}{2}-i\frac{p}{\cQ}
-\frac{q}{\cQ}+ \frac{\nu(\sigma)}{2}\right)} \ , \nn
&& \hspace{3cm} \nu(\NS)=0~,~~~ \nu(\R)=1~.
\label{hairpin bw}
\end{eqnarray}
This is naturally regarded as the supersymmetric extension of 
the ``hairpin brane'' presented in \cite{LVZ}.



To clarify the shape of D-brane \eqn{hairpin brane}
it is helpful to analyse the position space wave function
obtained by a simple Fourier analysis. 
Our goal is to derive the hairpin shape of D-brane from 
the boundary wave function \eqn{hairpin bw};
\begin{equation}
\tilde r e^{-\f{\cQ \phi}{2}}=2\cos \f{\cQ x}{2}~.
\label{hairpin}
\end{equation}
We focus only on the NS sector and fix $P=Q=0$ for simplicity.

The Fourier transform  of  \eqn{hairpin bw} is defined as 
\begin{equation}
\tilde{\Psi}^{(\sNS)}(\phi,x)\equiv \int_{-\infty}^{\infty} \f{dpdq}{(2\pi)^2} \,
e^{-ip\phi}e^{-iqx}\Psi^{(\sNS)}(p,q).
\end{equation}
Using the formula
\begin{equation}
\int_{-\infty}^\infty \f{e^{iat}dt}{\Gamma(x+t)\Gamma(y-t)}
= \theta(\pi -|a|)\f{[2\cos(a/2)]^{x+y-2}}{\Gamma(x+y-1)}e^{ia(y-x)/2},
\end{equation}
we can perform the $q$-integration  as follows
\begin{equation}
\int_{-\infty}^{\infty} \f{e^{-iqx}dq }{\Gamma(\f{1}{2}-\f{q}{\cQ}-i\f{p}{\cQ})
\Gamma(\f{1}{2}+\f{q}{\cQ}-i\f{p}{\cQ})}
=\cQ \f{\Big(2\cos\f{\cQ x}{2}\Big)^{-i\f{2p}{\cQ}-1}}
{\Gamma(-i\f{2p}{\cQ})}\, \theta(\f{\pi}{2}-\Big|\f{\cQ x}{2}\Big|).
\label{heviside}
\end{equation}
Note that this integral vanishes for
$|\f{\cQ x}{2}|>\f{\pi}{2}$ due to the Heaviside function.
This is consistent with the fact that the hairpin D-brane 
(\ref{hairpin}) lies only in the region $|\f{\cQ x}{2}|<\f{\pi}{2}$.
We will assume $|\f{\cQ x}{2}|<\f{\pi}{2}$ from now on.
After the integration  of $q$, 
the Fourier transformed wave function is given by
\begin{equation}
\tilde \Psi^{(\sNS)}(\phi,x)=
\f{\sqrt{2}}{\pi}\int_{-\infty}^{\infty} \f{dp}{2\pi}
\Big(2\cos\f{\cQ x}{2}\Big)^{-i\frac{2p}{\cQ}-1}e^{-ip\phi}
\Gamma(1-i\cQ p).
\label{calc1}
\end{equation}
%
%
Again we can perform the integral using the following formula
\begin{equation}
\int_{-i\infty}^{i\infty}dt \Gamma(a-t)z^t =2\pi i z^a e^{-z}\quad
\mbox{for} \quad a>0,
\end{equation}
yielding
\begin{equation}
\tilde \Psi^{(\sNS)}(\phi,x)=\f{\sqrt{2}}
{\pi\cQ \Big(2\cos \f{\cQ x}{2}\Big)^{\f{2}{\cQ^2}+1}}
\exp\Big[-\f{\phi}{\cQ}-\f{e^{-\f{\phi}{\cQ}}}
{\Big(2\cos \f{\cQ x}{2}\Big)^{\frac{2}{\cQ^2}}}\Big],
\label{bxy}
\end{equation}
%
We find the wave function $\tilde \Psi^{(\sNS)}(\phi,x)$ has a peak
at the trajectory of the hairpin curve (\ref{hairpin})
in the case of $\tilde{r}=1$, {\em i.e.}
$e^{-\f{\cQ \phi}{2}}=2\cos \f{\cQ x}{2}$.
Especially in the classical limit $N \,\sim \, \infty$, 
we can neglect the gamma function 
in eq.(\ref{calc1}) and $\tilde \Psi^{(\sNS)}(\phi,x)$ behaves
like the following delta function 
\begin{equation}
\tilde \Psi^{(\sNS)}(\phi,x) \sim 
\delta(\phi+\f{2}{\cQ}\ln 2\cos \f{\cQ x}{2})\sim
\delta(e^{-\f{\cQ \phi}{2}}-2\cos \f{\cQ x}{2})~,
\end{equation}
which reproduces the expected hairpin shape \eqn{hairpin}
with $\tilde{r}=1$.

It is easy to generalize to the cases with general $\tilde{r} >0$.
This is simply achieved by the zero-mode shift of $\phi$ as 
$\phi\, \rightarrow\, \phi' \equiv \phi - \frac{2}{\cQ} \ln \tilde{r}$
in \eqn{bxy}, which amounts to including the extra phase factor 
$e^{i\frac{2p}{\cQ} \ln \tilde{r}}$ in the momentum space wave function
\eqn{hairpin bw}. According to the standard argument in Liouville
theories, this phase factor could be identified as the contribution 
from the (renormalized) cosmological constant, 
as discussed in \cite{LVZ}, after incorporating the suitable 
Liouville potential which we will discuss later. 
One can also analyse likewise the cases with general $P$, $Q$. 
The inclusion of $Q$ simply gives rise to the parallel transport of 
the hairpin brane in the $X$-direction.\footnote
  {By the same reason the position space wave function in the R-sector
   is given by simply multiplying $e^{ix \cQ/2}$ to that of the NS sector.}
On the other hand, the inclusion of $P$ modifies 
the shape of hairpin in a more non-trivial manner 
along the radial direction $\phi$, although we 
do not give detailed calculations here. 


As opposed to \eqn{BPS brane}, the boundary states 
\eqn{hairpin brane} does not preserve the space-time SUSY at all.  
Namely, this is a non-BPS brane.\footnote
  {The non-BPS nature of \eqn{hairpin brane} originates from 
   the fact that, although it has been constructed based on an $\cN=2$ 
   superconformal structure, the GSO condition here is {\em not\/}
   correlated to the $U(1)$-current of this $\cN=2$ SCA. (See the 
   comments below.) Interestingly, it seems to be a different 
   feature from the familiar non-BPS branes in flat backgrounds 
   (see {\em e.g.} \cite{Sen-review}). In fact, our non-BPS brane 
   \eqn{hairpin brane} has non-vanishing RR-charges like 
   the usual BPS branes and is called ``BPS'' in \cite{Kutasov}.}  
To show this fact let us evaluate the self-overlap as before.
(We again assume $P=Q=0$ for simplicity.)
\begin{eqnarray}
Z &=& 
\int_0^\infty dp \int_{-\infty}^{\infty} dq \, \left\lb  
\rho^{(\sNS)}(p,q) \ch{(\sNS)}{}(p,q;it,0) \right. \cr
& & \left. - \rho^{(\stNS)}(p,q) \ch{(\stNS)}{}(p,q;it,0)+
\rho^{(\sR)}(p,q)\ch{(\sR)}{}(p,q;it,0)  \right\rb \ ,
\end{eqnarray}
where the density of states is given by
\begin{eqnarray}
\rho^{(\sNS)}(p,q) = \rho^{(\sR)}(p,q) &=&
\int_0^\infty dp' \int_{-\infty}^{\infty} dq' 
2\cos(2\pi pp')\cos(2\pi qq')| \Psi^{(\sNS)}(p',q')|^2 \cr
  &=& \rho_1(p) \delta(q) + \rho_2(p)[\delta(q-{\cQ}^{-1}) 
+ \delta(q+{\cQ}^{-1})] \ ,
\end{eqnarray}
for the open NS, R sectors, and 
\begin{eqnarray}
\rho^{(\stNS)}(p,q) &=&\int_0^\infty dp' 
\int_{-\infty}^{\infty} dq' 2\cos(2\pi pp')\cos(2\pi qq')
|\Psi^{(\sR)}(p',q')|^2 \cr
  &=& \rho_1(p) \delta(q) - \rho_2(p)[\delta(q-{\cQ}^{-1}) 
+ \delta(q+{\cQ}^{-1})] \ ,
\label{rho stNS}
\end{eqnarray}
for the $\tNS$ sector. 
The spectral density $\rho_i$ for each GSO projected sector 
are given in \eqn{rho 1 0}, \eqn{rho 2 0}.

These expressions clearly explain the aspect of open GSO projection. 
The sign difference in front of the $\rho_2$ originates again from 
the remark around \eqn{identity 1}. It indicates that 
the string stretching between the different asymptotic sides 
of hairpin have the opposite GSO projection from 
that on the same side. 
This fact implies that SUSY cancellation occurs only in the 
$\rho_1$-part, while does not for the $\rho_2$-part.

On the other hand, let us recall the BPS brane case \eqn{BPS brane}. 
Thanks to the interplay between the spectral flow summation 
and the minimal model characters, the net contribution 
from this sign difference cancels out.  
In the end we have the same SUSY cancellation mechanism for 
both the $\rho_1$ and $\rho_2$ parts.

To close this section we make a few comments:

~


\noindent
{\bf 1. } As promised before, we here discuss the proper marginal
deformation to avoid the singularity at $\phi \, \sim \, -\infty$.
One might suppose that we only have to use the chiral 
Liouville potential \eqn{cosm term 1} with simply replacing $Y$
with $X$. However, this is found not to work. In fact, the space-time SUSY
in the CHS background is realized by the spin fields defined by 
the bosonizations as 
\begin{eqnarray}
 i \partial H_0 = \psi^0 \psi^1~, ~~
 i \partial H_1 = i\psi^2 \psi^3~,~~
 i \partial H_2 = i\psi^4 \psi^5~, ~~
 i \partial H_3 = i\chi^1 \chi^2~, ~~
 i \partial H_4 = i\chi^3 \psi^{\phi}~,
\label{spin fields}
\end{eqnarray} 
where $\psi^{\mu}$, ($\mu=0,\ldots, 5$), 
$\psi^{\phi}$, $\chi^i$, ($i=1,2,3$) are free fermions 
along the directions $\br^{5,1}$, $\br_{\phi}$, $SU(2)$ 
respectively. (We can choose, say, $\psi^1= \psi^X$.)
The requirement of BRST invariance makes the $H_3$ and $H_4$ spin 
fields correlate, leaving only the  16 supercharges 
(together with the right mover) as is well-known. 
In other words the space-time supercharges must include 
the spectral flow operators associated to the total $\cN=2$ 
$U(1)$-charge in  the coupled system:
$\dsp \f{M_{N-2} \times \lb \br_{\phi} \times S^1_Y \rb}{\bz_N}$. 
The Liouville potential \eqn{cosm term 1} is compatible with the 
space-time SUSY in this sense, while it is {\em not\/} 
after replacing $Y$ with $X$. 
This type Liouville potential is not local 
with the supercharges constructed from \eqn{spin fields} and  
breaks the translational invariance in $\br^{5,1}$.  
Consequently, we instead propose  to take a marginal deformation 
of the ``non-chiral Liouville potential'' (up to total derivative);
\begin{eqnarray}
\mu ' \int d^2z \, 
\left(i\partial X + i \cQ \psi^{\phi} \psi^X\right) 
\left(i\deebar X + i \cQ \tilde{\psi}^{\phi} \tilde{\psi}^X\right)
e^{-\cQ \phi}~.
\label{cosm term 2} 
\end{eqnarray}
This preserves the $\cN=2$ superconformal symmetry, and furthermore
is compatible with the GSO projection, 
since it includes {\em even\/} number of fermions.  
It also preserves the translational invariance because of the absence of 
zero-mode of $X$.
We note that the boundary wave function \eqn{hairpin bw}
is the one consistent with the $\cN=2$ 
Liouville interactions, which is obvious from the construction.
Especially, it satisfies the correct reflection relation \cite{BF,AKRS}. 
We thus conclude that our hairpin brane \eqn{hairpin brane}
is consistent even under the marginal deformation \eqn{cosm term 2}.
It is also worthwhile to mention on the duality 
between the marginal deformations of the types \eqn{cosm term 2} and 
\eqn{cosm term 1} (with $Y$ replaced by $X$) conjectured in 
\cite{AKRS} (see also \cite{Nakayama,IPT2}).

~

\begin{figure}[h!]
\begin{center}
\scalebox{0.6}{\includegraphics{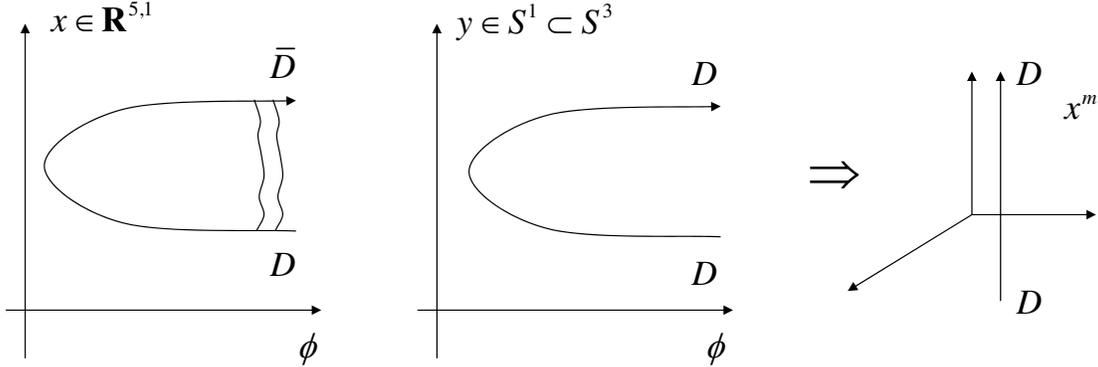}} \hspace{5mm}
\caption{Hairpin curves in several directions
\label{fig:hairpin}}
\end{center}
\end{figure}

\noindent
{\bf 2. } 
As we discussed above, the hairpin brane \eqn{hairpin brane} 
is non-BPS.
The most intuitive understanding of this fact is 
achieved by observing the Chern-Simons term in the DBI action for 
the hairpin brane. By directly substituting the hairpin solution
\eqn{hairpin}, one can easily find that the hairpin brane 
\eqn{hairpin brane} looks like $D\bar{D}$-system in the asymptotic 
region $\phi \, \sim \, + \infty$. 
This fact matches with the above analysis 
of cylinder amplitudes, supporting the validity of the boundary 
wave function \eqn{hairpin bw}. 
The $\rho_2$-part of amplitude is identified 
as the $D\bar{D}$-open strings in the asymptotic region. 
It is worth noting that the tachyonic modes in such $D\bar{D}$-open strings 
are actually massive for arbitrary $N \geq 2$ due to the energies 
of stretched strings with the length 
$2\pi/\cQ = \pi\sqrt{2N} $ (together with the massgap $\cQ^2/8= 1/(4N)$). 
See the left figure in fig.\ref{fig:hairpin}.
Therefore, this brane is a non-trivial example of stable non-BPS D-branes.
\footnote{Here, ``stable'' means nothing but to have no tachyonic
modes in the open string channel. However, it is plausible to expect
that this brane is really stable, at least perturbatively,
against the gravitational attractive force after turning on
the Liouville interaction, since we could not have any marginal
perturbation consistent in the boundary $\cN=2$ Liouville theory
which alters the distance $\sim 1/\cQ$ between the asymptotic
$D$-$\bar{D}$ branes. It is challenging and interesting to
investigate this aspect more rigorously.}
Note that the Liouville potential \eqn{cosm term 2} could prevent 
the stretched $D\bar{D}$-open strings from shrinking around 
the corner of hairpin.

On the other hand, the BPS brane \eqn{BPS brane} looks like 
$DD$-system rather than $D \bar{D}$ in the asymptotic region.
To figure out this fact,  
let us first recall the CHS geometry (in the string frame);
\begin{equation}
\begin{split}
ds^2&=\eta_{\mu\nu}dx^\mu dx^\nu +\Big(1+\f{N\alpha'}{r^2}\Big)
dx^m dx^m,\quad e^{2(\Phi-\Phi_0)}=1+\f{N\alpha'}{r^2},\quad
H_{mnp}=-\epsilon^q_{mnp}\partial_q \Phi
\end{split}\label{CHS bg}
\end{equation}
where we defined $r^2=x^m x^m$ and the linear dilaton coordinate is 
related as $\phi = \sqrt{N\alpha' } \ln \sqrt{\f{r^2}{N\alpha'}}$
in the near horizon limit.
The index $\mu$ runs 0 to 5 and the index $m$
runs 6 to 9.
Consider the hairpin D-brane whose $x$ direction $(S^1)$ is
chosen in $S^3$ (see the right figure of fig.\ref{fig:hairpin}).
At first glance it may appear asymptotically
a $D\bar D$ system as above,
since the direction of the D-brane becomes opposite 
in the asymptotic region in this figure.
However this is not the case, because
this hairpin curve (\ref{hairpin}) is a
straight line in the coordinates $x^m$.
In the CHS background with the coordinates
(\ref{CHS bg}),  Killing spinors are constant up
to a certain function $f(r)$, implying 
this is indeed a $DD$ system. In other words, the two asymptotic 
sides of hairpin correspond respectively 
to the D0-branes located at the north pole
({\em i.e.} $\ket{L=0}$) and the south pole ($\ket{L=N-2}$) 
of $S^3$,  both of which have the same signature 
of the RR-charge. 
All of these features completely match with our observations 
of the cylinder amplitudes, especially with 
respect to the sign difference in the $\rho_2$-terms
between \eqn{self-overlap 2} and \eqn{rho stNS}.


~

\section{Rolling D-branes in NS5 Backgrounds}

~

As we declared, we now attempt to make the Wick rotation of 
$X$ to the time coordinate $X^0$ in the non-BPS hairpin brane 
\eqn{hairpin brane} to construct the boundary states describing
the rolling D-brane solution (accompanied by the replacements
$X\,\rightarrow\,X^0$, $\psi^X\,\rightarrow\,\psi^{X^0}$ in 
\eqn{cosm term 2}). 
We again concentrate on the $P=Q=0$ case
to avoid unessential complexities. 

~

\subsection{Wick Rotation to the Rolling D-brane}

~

First of all,  let us try to make 
the naive Wick rotation $q \rightarrow i\omega$ in \eqn{hairpin bw},
{\em i.e.}
\begin{equation}
\Psi^{(\sNS)}_{\msc{naive}}
(p,\omega)=\sqrt{2}\cQ e^{i\f{2 p}{\cQ}\ln \tilde r }
\f{\Gamma(-i\cQ p)\Gamma(1-i\f{2p}{\cQ})}
{\Gamma(\f{1}{2}+i\f{\omega}{\cQ}-i\f{p}{\cQ})
\Gamma(\f{1}{2}-i\f{\omega}{\cQ}-i\f{p}{\cQ})}~,
\label{bpe naive}
\end{equation}
where we introduced the parameter $\tilde{r}$ in \eqn{hairpin}
which will be identified as $\tilde r=\f{2E}{\tau_p}$ in 
the rolling D-brane solution \eqn{rolling solution}.
Unfortunately this result is physically unacceptable because
of the bad UV behavior. 
We cannot Fourier transform it and thus cannot reproduce 
the correct trajectory of the rolling D-brane
solution \eqn{rolling solution}. 
In fact, the absolute square of this wave function behaves as 
\begin{equation}
|\Psi_{\msc{naive}}^{(\sNS)}(p,\omega)|^2 
\sim \f{\cosh(\f{2\pi \omega}{\cQ})+\cosh(\f{2\pi p}{\cQ})}
{\sinh(\pi \cQ p)\sinh(\f{2\pi p}{\cQ} )}~,
\end{equation}
which is divergent under $\omega \rightarrow \infty$.

On the other hand,  the following wave function 
is physically acceptable, which is obtained 
via the Wick rotation in the {\it position} space $x \rightarrow it$;
\begin{equation}
\tilde \Psi^{(\sNS)}(\phi,t)=\f{\sqrt{2}}
{\pi{\cQ}\Big(2 \cosh \f{\cQ t}{2}\Big)^{\frac{2}{\cQ^2}+1}}
\exp\Big[-\f{\phi'}{\cQ}-\f{e^{-\frac{\phi'}{\cQ}}}
{\Big(2 \cosh \f{\cQ t}{2}\Big)^{\f{2}{\cQ^2}}}\Big]
,\quad \phi' \equiv \phi-\f{2}{\cQ}\ln \tilde r.
\label{bxt}
\end{equation}
As is obvious from the construction,
the peak of this wave function is located at the trajectory
of the rolling D-brane (\ref{rolling solution}).
In addition this wave function vanishes as $|t| \rightarrow \infty$
due to the suppression factor $\sim 1/\cosh t$,
as expected from the low energy analysis using the DBI action.
It has a well-defined Fourier transform;
\begin{equation}
\begin{split}
\Psi^{(\sNS)}(p,\omega) &=\int_{-\infty}^\infty d\phi\int_{-\infty}^\infty
dt \, e^{ip\phi}e^{-i\omega t}\tilde{\Psi}^{(\sNS)}(\phi,t)\\
&=\f{\sqrt{2}}{\pi} e^{\f{2ip}{\cQ}\ln \tilde{r}} \Gamma(1-i \cQ p)
\int_{-\infty}^\infty dt \,
\Big(2\cosh\f{\cQ t}{2}\Big)^{-i\f{2p}{\cQ}-1}e^{-i\omega t}.
\end{split}
\end{equation}
In fact, the integral in the second line is manifestly convergent.

One might ask why the naive Wick-rotation leads to a wrong result. 
The answer is very simple: Due to the existence of Heaviside function 
the R.H.S. of the formula \eqn{heviside} is not analytic 
and we cannot analytically continue it.

\begin{figure}[h!]
\begin{center}
\scalebox{0.7}{\includegraphics{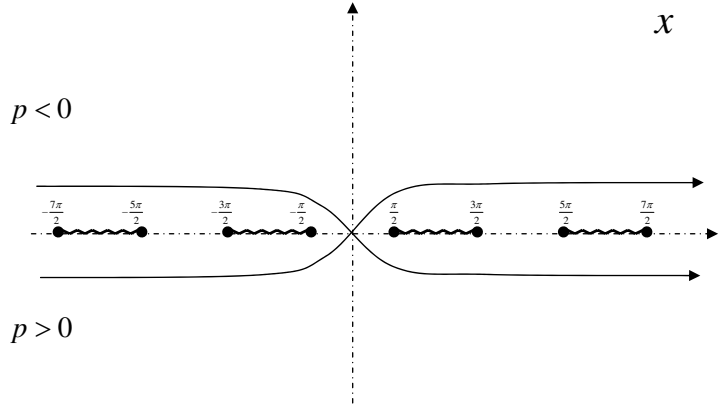}} \hspace{5mm}
\caption{Branch cuts and contour
\label{fig:contour}}
\end{center}
\end{figure}

Another natural question would be how we can reproduce the position space 
wave function \eqn{bxt} from the one in the momentum space.  
To answer the question we return to the wave function 
of hairpin brane \eqn{hairpin bw}.
To simplify the arguments we eliminate $\cQ$ 
by redefining the momenta and positions
like $\frac{2p}{\cQ} \rightarrow p$ and $\frac{\cQ x}{2} \rightarrow x$. 
Again due to this Heaviside function, the contribution of
the hairpin D-brane comes only from the region $-\pi/2<x<\pi/2$,
\begin{equation}
\int_{-\f{\pi}{2}}^\f{\pi}{2}dx\,(2\cos x)^{-ip-1} e^{iqx}
=\f{\pi \Gamma(-ip)}
{\,\Gamma(\f{1}{2}-\f{q}{2}-i\f{p}{2})\Gamma(\f{1}{2}+\f{q}{2}-i\f{p}{2})}.
\end{equation}
Then we need to extend the integration contour to an infinite line
to make the Wick rotation possible with keeping the integral 
convergent.  
Note that the function $\ln \cos x$ has
branch cuts between $-\f{3\pi}{2}+2\pi m<x< -\f{\pi}{2}+2\pi m$, 
($m \in \bz$) on the real line. We so  find that the contour should be set
as in fig.\ref{fig:contour}.

For $p>0$ the integral can be calculated as follows
\begin{equation}
\begin{split}
&\int_{\cC} dx\,[2\cos x]^{-ip-1}e^{iqx}\\
&=\lim_{\epsilon \to 0^+}\int_{-\infty}^\f{\pi}{2}
dx\,[2\cos(x-i\epsilon)]^{-ip-1}e^{iqx}
+\lim_{\epsilon \to 0^+}\int_{\f{\pi}{2}}^\infty
dx\,[2\cos(x+i\epsilon)]^{-ip-1}e^{iqx}\\
&=\Big[\sum_{m=-\infty}^0(-e^{\pi p}e^{i\pi q})^m
+\sum_{m=1}^\infty(-e^{-\pi p}e^{i\pi q})^m
\Big]\int_{-\f{\pi}{2}}^\f{\pi}{2}dx\,(2\cos x)^{-ip-1} e^{iqx}\\
&=\Big[\f{1}{1+e^{-\pi(p+iq)}}+\f{1}{1+e^{-\pi(p-iq)}}-1\Big]
\f{\pi \Gamma(-ip)}
{\Gamma(\f{1}{2}-\f{q}{2}-i\f{p}{2})\Gamma(\f{1}{2}+\f{q}{2}-i\f{p}{2})}.
\end{split}\label{bpq array0}
\end{equation}
The third line in (\ref{bpq array0}) takes the form
of the summation over the contribution of each hairpin brane 
located in the interval 
$\f{2m-1}{2}\pi< x < \f{2m+1}{2}\pi$.
We can likewise calculate for $p<0$ and the result is summarized 
as follows;
\begin{equation}
\begin{split}
&\int_{\cC} dx\,[2\cos x]^{-ip-1}e^{iqx}\\
&=\f{\sgn(p)\sinh(\pi p)}{2\cosh[\f{\pi}{2}(p+iq)]\cosh[\f{\pi}{2}(p-iq)]}
\times \f{\pi \Gamma(-ip)}
{\Gamma(\f{1}{2}-\f{q}{2}-i\f{p}{2})\Gamma(\f{1}{2}+\f{q}{2}-i\f{p}{2})}.
\end{split}
\label{bpq array}
\end{equation}
Note that the additional factor is an even function of $p$.
Hence the derived  wave function is
compatible with the reflection relation in  the $\cN=2$ Liouville
theory.

Now, for $-1<q<1$,  we can Wick-rotate the contour $\cC$
to the imaginary line $(x=it)$ with avoiding crossing the branch cuts;
\begin{equation}
\begin{split}
&\int_{-i\infty}^{i\infty} dx\,[2\cos x]^{-ip-1}e^{iqx}
=i\int_{-\infty}^\infty dt\,[2\cosh t]^{-ip-1}e^{-qt}\\
&=
\f{\sinh(\pi p)}{2\cosh[\f{\pi}{2}(p+iq)]\cosh[\f{\pi}{2}(p-iq)]}
\f{\pi \Gamma(-ip)}
{\Gamma(\f{1}{2}-\f{q}{2}-i\f{p}{2})\Gamma(\f{1}{2}+\f{q}{2}-i\f{p}{2})}.
\label{BPE0}
\end{split}
\end{equation}
Since both sides of ({\ref{BPE0}}) are clearly analytic
with respect to $q$ around the imaginary line,
the analytic continuation $q=i\omega$ can be performed. 
This continuation leads to
  \footnote{We have also checked the formula (\ref{BPE})
    numerically using MATHEMATICA.}
\begin{equation}
\begin{split}
&\int_{-\infty}^{\infty}dt\,[2\cosh t]^{-ip-1}e^{-i\omega t}\\
&=\Big[\f{1}{1+e^{-\pi(p+\omega)}}+\f{1}{1+e^{-\pi(p-\omega)}}-1\Big]\times
\f{-i\pi \Gamma(-ip)}
{\Gamma[\f{1}{2}-\f{i}{2}(p-\omega)]\Gamma[\f{1}{2}-\f{i}{2}(p+\omega)]}\\
&=\f{\sinh(\pi p)}{2\cosh[\f{\pi}{2}(p+\omega)]\cosh[\f{\pi}{2}(p-\omega)]}
\times \f{-i\pi \Gamma(-ip)}
{\Gamma[\f{1}{2}-\f{i}{2}(p-\omega)]\Gamma[\f{1}{2}-\f{i}{2}(p+\omega)]}.
\end{split}
\label{BPE}
\end{equation}
It is remarkable that we now have an additional suppression factor
which improves the UV behavior of $\omega$.
This is quite reminiscent of the one characteristic for 
the rolling tachyon solution (full s-brane) \cite{Sen-RT}.


Finally, after restoring $\cQ$, we find that 
the rolling D-brane can be expressed in the momentum
space as follows;
\begin{eqnarray}
\Psi^{(\sNS)}(p,\omega)&=&\f{-i\sinh(\f{ 2 \pi p}{\cQ})}
{2\cosh[\f{\pi}{\cQ}(p+\omega)]\cosh[\f{\pi}{\cQ}(p-\omega)]} \,
\Psi^{(\sNS)}_{\msc{naive}}(p,\omega) \nn
&=& \f{-i \sqrt{2}\cQ e^{i\f{2 p}{\cQ}\ln \tilde r }\sinh(\f{ 2 \pi p}{\cQ})}
{2\cosh[\f{\pi}{\cQ}(p+\omega)]\cosh[\f{\pi}{\cQ}(p-\omega)]} \cdot
\f{\Gamma(-i\cQ p)\Gamma(1-i\f{2p}{\cQ})}
{\Gamma(\f{1}{2}+i\f{\omega}{\cQ}-i\f{p}{\cQ})
\Gamma(\f{1}{2}-i\f{\omega}{\cQ}-i\f{p}{\cQ})}.
\label{NSwick}
\end{eqnarray}
Its absolute square is 
\begin{eqnarray}
|\Psi^{(\sNS)}(p,\omega)|^2 = 
\frac{2\sinh\left(\frac{2\pi p}{\cQ}\right)}
{\left[ \cosh\left(\frac{2\pi \omega}{\cQ}\right) 
+ \cosh\left(\frac{2\pi p}{\cQ}\right)\right]\sinh(\pi \cQ p)} \ .
\end{eqnarray}

In the R sector the Wick rotation becomes a little bit different. 
If we substitute $q = i\omega$ into the R wave function 
after the Wick rotation as we have just done, we obtain
\begin{eqnarray}
\Psi^{(\sR)}(p,\omega) 
=\frac{-i \sqrt{2}\cQ e^{i\f{2 p}{\cQ}\ln \tilde r }\sinh\left(\frac{2\pi p}{\cQ}\right)}
{\cosh\left(\frac{2\pi p}{\cQ}\right) - 
\cosh\left(\frac{2\pi \omega}{\cQ}\right)} \cdot
\frac{\Gamma(-i\cQ p)\Gamma\left(1-i\frac{2p}{\cQ}\right)}
{\Gamma\left(1-i\frac{p}{\cQ}-i\frac{\omega}{\cQ}\right)
\Gamma\left(-i\frac{p}{\cQ}+i\frac{\omega}{\cQ}\right)} \ .
\label{Rwick}
\end{eqnarray}
One of the easiest way to obtain this result 
is to substitute $\omega_{\sNS} = \omega_{\sR} -\frac{i\cQ}{2}$ 
into the NS wave function (\ref{NSwick}). Its absolute square is 
given by
\begin{eqnarray}
|\Psi^{(\sR)}(p,\omega)|^2 =  
\frac{\omega-p}{\omega+p}\frac{2\sinh\left(\frac{2\pi p}{\cQ}\right)}
{\left[\cosh\left(\frac{2\pi \omega}{\cQ}\right) 
- \cosh\left(\frac{2\pi p}{\cQ}\right)\right]\sinh(\pi \cQ p)}\ .
\end{eqnarray}
The subtlety here is that we no longer have a sensible GSO projection 
for the open sector. Recalling the symmetry 
with respect to $p$ and $-p$, we can rewrite this 
in a more suggestive form 
\begin{eqnarray}
|\Psi^{(\sR)}(p,\omega)|^2 =  
\frac{\omega^2+p^2}{(\omega^2-p^2)}
\frac{2\sinh\left(\frac{2\pi p}{\cQ}\right)}
{\left[\cosh\left(\frac{2\pi \omega}{\cQ}\right) 
- \cosh\left(\frac{2\pi p}{\cQ}\right)\right]\sinh(\pi \cQ p)}\ .
\end{eqnarray}
It is interesting to observe that the R sector 
has a singularity when $p=\pm \omega$ {\em i.e.} 
in the case of the massless forward (and backward)
emission from the rolling D-brane.

Several comments are in order:

~

\noindent
{\bf 1.} The  ``time-like $\cN=2$ Liouville theory'' we are considering
here is defined only by the Wick rotation of $X$ 
with leaving $\phi$ space-like, and not accompanied 
by the analytic continuation of any parameter in the model
(say, the background charge $\cQ$). This point is in a sharp
contrast with the bosonic time-like Liouville theory 
\cite{GS2,ST,Schomerus}, in which the time-like $\phi$ is considered 
and also the analytic continuation of the parameter is taken.

~

\noindent
{\bf 2.} As we already addressed,
the boundary wave function \eqn{NSwick}
for the rolling D-brane we proposed 
includes a suppression factor for energy $\om$, which improves 
the UV behavior and reproduces the correct profile of rolling brane. 
Although this seems quite satisfactory, 
a subtlety now appears since this factor is an odd function of $p$.
One may be afraid that it would contradict 
the reflection amplitude in the Liouville
theory. However, it originates from the Wick rotation, namely,
the factor $\sgn(p)$ in eq.(\ref{bpq array}) is canceled
by the orientation of the Wick-rotated contour. 
We again stress that our boundary wave function 
is originally consistent with the reflection amplitude  
before performing the Wick rotation (recall \eqn{bpq array}), 
and thus should be the correct one 
according to the spirit of Wick rotation.   
It may be interesting to further investigate whether 
such peculiarity is a general feature in the time-like $\cN=2$ Liouville
theory.

~

\noindent
{\bf 3. } 
The prefactor discussed above has the similar form 
to that for the rolling tachyon solution, as already mentioned.
This fact reminds us of the interpretation of the rolling tachyon 
solution (with a certain coupling constant)
as an array of the sD-branes on the imaginary time axis 
discussed in \cite{Sen-RT,MSY,LLM,GIR}. 
It may be fascinating to ask whether the similar interpretation 
is possible for the rolling D-brane: 
the interpretation as an array of the infinite hairpin branes  
along the imaginary axis.

~

\noindent
{\bf 4.}
Another natural question is whether
the rolling brane boundary state \eqn{NSwick} really satisfies 
the Cardy condition. It is not so difficult to confirm 
this is indeed the case among the rolling branes themselves,
as well as with arbitrary static branes that are trivial along the 
$\phi$, $X^0$-directions. However, if we consider 
the branes associated to {\em degenerate\/} 
representations in the time-like $\cN=2$ Liouville theory here, 
the situation gets quite subtle. 
For example, one might consider the ``time-dependent class 1 brane''
(ZZ-type) associated to the identity representation \cite{ES-L,ASY}. 
The overlap with it should satisfy the modular bootstrap equation. 
Unfortunately, it does not seem to be the case, 
since the prefactor considered here is not likely to be identified 
as a modular coefficient of any unitary representation of $\cN=2$ SCA.
However, we here note that the {\em time-like\/} characters 
of degenerate representations cannot completely cancel 
the contributions from the ghost oscillators 
due to the existence of singular vectors. 
From this fact, one can easily find that any time-dependent 
brane of a degenerate representation always include 
negative norm states in the open channel, 
{\em even if it is a unitary representation.} Hence,
it is plausible to discard such time-dependent ``degenerate''  branes,  
and in this sense,  our rolling brane \eqn{NSwick} 
does not contradict the Cardy condition. 
Further studies would be needed for this issue.

~


\subsection{Open String Density of States and Closed String Emission}

~

By using the boundary wave functions we obtained \eqn{NSwick}, \eqn{Rwick},
we can evaluate 
the open channel density of states and the closed string emission rate
in the rolling process.
The density of states in the open channel is calculated in the same 
way as the previous analysis\footnote
  {The Gaussian integral of zero-mode along the 
   time direction is apparently divergent due to its wrong sign in 
   the exponent. Therefore, to perform such a modular transformation, 
   we shall implicitly define it as an analytic continuation 
   of the Euclidean calculation, 
   or assume the Lorentzian signature worldsheet.}
\begin{eqnarray}
\rho^{(\sNS)} (p',\omega') &=&  \int_{-\infty}^{\infty} 
d\omega \int_0^{\infty}dp\, 2 \cos(2\pi pp') \cos(2\pi \omega \omega') 
\frac{2\sinh\left(\frac{2\pi p}{\cQ}\right)}
{\left[ \cosh\left(\frac{2\pi \omega}{\cQ}\right) 
+ \cosh\left(\frac{2\pi p}{\cQ}\right)\right]\sinh(\pi \cQ p)} \cr
 &=&  \frac{2\sinh\left(\frac{2\pi \omega'}{\cQ}\right)}
{\left[ \cosh\left(\frac{2\pi \omega'}{\cQ}\right) 
+ \cosh\left(\frac{2\pi p'}{\cQ}\right)\right]\sinh(\pi \cQ \omega')} \ ,
\end{eqnarray}
for the NS sector by using the Fourier transformation formula;
\begin{eqnarray}
\int_{-\infty}^{\infty} \f{dk}{2\pi} 
\frac{\sinh(a\pi)}{\cosh(k)+\cosh(a\pi)} 
e^{ikx} = \frac{\sin(a\pi x)}{\sinh(\pi x)} \ . 
\end{eqnarray}
It is amazing that the result is simply given 
by the exchange of momentum and energy of the \textit{closed} channel 
amplitudes. This self-dual property of the amplitudes 
is peculiar to this function. 
At the same time, the result shows that our open channel density 
is manifestly positive definite as it should be.

On the other hand, for the R sector or the open $\tNS$ sector equivalently, 
the density of states has a divergence 
on the light-cone direction $\omega = \pm p$.
\begin{eqnarray}
&&\rho^{(\stNS)} (p',\omega') = \int_{-\infty}^{\infty} 
d\omega \int_0^{\infty}dp\, 
2 \cos(2\pi pp') \cos(2\pi \omega \omega') \nn
&& \hspace{3cm} \times
\frac{\omega^2+p^2}{(\omega^2-p^2)}
\frac{2\sinh\left(\frac{2\pi p}{\cQ}\right)}
{\left[\cosh\left(\frac{2\pi \omega}{\cQ}\right) 
- \cosh\left(\frac{2\pi p}{\cQ}\right)\right]\sinh(\pi \cQ p)}\ . 
\end{eqnarray}
This divergence comes from the double pole 
of the absolute square of the boundary wave function, 
which is typical of the infrared divergence 
for the extending branes in the Liouville theories (FZZT-like branes). 
This fact suggests that our brane is extending 
along a non-compact direction and reaches 
the speed of light in the far past and future, 
as is expected. 
The reason why the NS amplitude, in contrast, does not 
have such a divergence is  intuitively understood as follows: 
In the CHS background the string coupling tends to be stronger and 
stronger as the brane comes close to the NS5-brane. 
As a result, the tension of the brane gets  smaller 
in the far past and future, and thus the gravitational interaction 
practically vanishes. On the other hand, the RR scattering 
does not damp because the coupling to the RR field does not depend on 
the value of the dilaton.  Hence, in view of the RR field probe, 
the rolling brane is actually extending from the past to the future. 

The difference between the density of states in the open NS 
and $\tNS$ sector obviously shows that our rolling brane 
is not supersymmetric nor GSO projected, as is expected since  we  
started from the non-BPS hairpin brane \eqn{hairpin bw}.



Now let us calculate the closed string emission rate. 
We again begin with the NS sector.
Analysing the self-overlap and using the optical theorem, 
the emission rate for a fixed transverse mass $M$ is 
evaluated in the similar manner to \cite{LLM,KLMS} as   
\begin{eqnarray}
\bar{N}^{(\sNS)}(M) = \int_{0}^{\infty} \frac{dp}{\omega_p} \,
\frac{2\sinh\left(\frac{2\pi p}{\cQ}\right)}
{\left[ \cosh\left(\frac{2\pi \omega_p}{\cQ}\right) 
+ \cosh\left(\frac{2\pi p}{\cQ}\right)\right]\sinh(\pi \cQ p)} \ , \ 
\omega_p = \sqrt{p^2 + M^2} \ .
\end{eqnarray}
It is convergent if $M^2 > 0$.
On the other hand, for the R sector, the emission rate is given by 
\begin{eqnarray}
\bar{N}^{(\sR)}(M) =  \int_{0}^{\infty} \frac{dp}{\omega_p} \,
\frac{\omega_p^2+p^2}{(\omega_p^2-p^2)}
\frac{2\sinh\left(\frac{2\pi p}{\cQ}\right)}
{\left[\cosh\left(\frac{2\pi \omega_p}{\cQ}\right) 
- \cosh\left(\frac{2\pi p}{\cQ}\right)\right]
\sinh(\pi \cQ p)}\  , \ \omega_p = \sqrt{p^2 + M^2} \ .
\end{eqnarray}
This  emission rate has a strong peak at 
the massless forward scattering $M=0, p=\omega$, 
but otherwise it is convergent.

Let us evaluate the asymptotic behaviors of these emission rates 
in the large $M$ limit. It is obvious that $\bar{N}^{(\sNS)}(M)$ and 
$\bar{N}^{(\sR)}(M)$ have the same asymptotic behaviors. 
We consider D$p$-brane, which has $d = 5-p$ transverse 
directions to the D-brane in $\br^5$.
Then the asymptotic emission rate 
behaves as 
\begin{eqnarray}
\bar{N}(M) \sim \int d^dk_{\perp}\int_0^{\infty} \frac{dp}{M}\, 
e^{\left(\frac{2\pi}{\cQ}-\pi \cQ\right)p - \frac{2\pi}{\cQ} \sqrt{p^2+k^2_\perp+M^2}}
\sim  M^{\frac{-1+d}{2}}e^{-2\pi M \sqrt{1-\frac{\cQ^2}{4}}}~,
\end{eqnarray}
where we used the saddle point approximation. 
The density of levels as a function of mass, 
however, is given by $ n(M) \propto M^{-3}e^{2\pi M \sqrt{1-\f{\cQ^2}{4}}}$, 
which is obtained by recalling $c_{\msc{eff}} = 12 - 3 \cQ^2$ 
because of the linear dilaton background \cite{KutS}.\footnote{
The boundary state couples only to the left-rignt symmetric closed string 
states, so the asymptotic density of states needed here is 
the square root of the full closed string density of 
states as is pointed out in \cite{Sahakyan}. 
}
Thus, the total emission rate becomes
\begin{eqnarray}
\frac{\bar{N}}{V} \sim \int^{\infty} dM\, M^{-\frac{p}{2}-1} ~, ~~~
\frac{\bar{E}}{V} \sim \int^{\infty} dM\, M^{-\frac{p}{2}} \ .
\end{eqnarray}
The total emitted energy has a powerlike divergence when $p\le 2$.
This fact contrasts with the UV finite brane decay 
in the bosonic linear dilaton background studied 
in \cite{KLMS}, and rather leads us to the same behavior 
as the decay of the non-BPS brane in the flat Minkowski space \cite{LLM}.
\footnote
    {In \cite{KLMS} the emission rate does not depend on 
     the background charge $\cQ$ while the effective number of 
     highly excited closed strings is reduced 
     by the linear dilaton background. 
    In our case, however, the emission rate compensates 
    the density of states, canceling precisely the $\cQ$-dependence.  
}


In this subsection we have discussed only the one closed string emission
from the rolling brane.   Owing to this limitation 
the rate of emission to the tangential direction 
to the brane vanishes simply by the momentum conservation. 
In the processes of two or more string emissions, 
the amplitudes for the tangential emissions could be non-zero.
However, from the experience in the two-dimensional string theories 
(see, {\em e.g.} \cite{MV,KMS,TT,DKKMMS,Gutperle:2003ij}), 
we believe that the inclusion of these decaying modes 
does not improve the qualitative properties of the worldsheet 
analysis. The quantum treatment of the D-brane should be
necessary in order to truly understand the fate of rolling D-brane.

~


\section{Summary and Discussions}

~

In this paper we studied a solution of time-dependent D-branes,
the rolling D-branes, in the NS5 background by means of the BCFT
approach. An interesting point addressed in \cite{Kutasov} 
is that the rolling D-brane has similarities 
to the rolling tachyon, 
which suggests that the BCFT analysis is helpful for further studies.
With this motivation, we have constructed the boundary state 
for the rolling D-brane.  
The Wick-rotated version of the rolling D-brane is
a hairpin brane whose boundary state has been constructed in the bosonic
theory in \cite{LVZ}. 
We can supersymmetrize the hairpin D-brane as the class 2 (FZZT)
D-brane in ${\cal N}=2$ Liouville theory \cite{ES-L,ASY}. 
This is quite natural because ${\cal N}=2$ Liouville theory 
is regarded as a supersymmetric extension of the sine-Liouville theory 
where the bosonic hairpin brane is considered in \cite{LVZ}.

These hairpin D-branes can be embedded in the CHS background in two ways.
When the direction of $\cN=2$ $U(1)$-current is  set to be in the internal
direction, {\em i.e.} $S^1$ in $S^3$, the hairpin D-brane is a BPS one. 
We confirmed this fact by analysing explicitly 
the self-overlap of boundary states.
On the other hand, when we choose the $U(1)$-direction parallel to 
NS5 (and decompactify it), 
the hairpin D-brane breaks all supersymmetries.
In this case we found the non-vanishing self-overlap which  
suggests that the two sides of hairpin looks like 
a $D\bar{D}$-system in the asymptotic region.  
Moreover, this is a stable non-BPS brane. 
This fact is physically understood as follows:
The Liouville interaction prevents the $D\bar{D}$ open strings 
from shrinking to the corner of the hairpin 
and the mass square of the tachyon modes actually gets positive.

As the highlight of this work, 
we have proposed an appropriate way of the Wick rotation to the 
rolling D-brane. 
By performing the Wick rotation in the {\it position} space, 
we have arrived at a sensible boundary wave function that provides
well-defined spectral densities of open string states. 
Furthermore, it successfully reproduces 
the correct trajectory of rolling D-brane 
under the classical limit.
It is remarkable that our procedure of Wick rotation
gives rise to the prefactor quite analogous to that for 
the rolling tachyon solution \cite{Sen-RT}, which improves the 
UV behavior of the boundary wave function \footnote{
Recently, an interesting prescription of
analytic continuation to time-like theory in the rolling tachyon
system has been discussed in \cite{BKKN} based on some matrix
integral techniques. That prescription also gives rise to non-trivial
suppression factors of energy that could improve the UV behaviors.
It would be an interesting problem to explore a relation to the way of
Wick rotation we made in this paper. We would like to thank
V. Balasubramanian for informing us of the paper \cite{BKKN}. }. 
This fact is likely to support the similarity to the rolling tachyon
recognized by Kutasov. 
This prefactor could be interpreted to originate from 
an infinite array of the Euclidean hairpin branes 
in the similar manner as the rolling tachyon case 
\cite{Sen-RT,MSY,LLM,GIR}.
We also calculated 
the closed string emission rate in  both the NS and R sectors. 
The different behaviors of emission rates between the NS and R sectors
have been found and explained from the physical viewpoints.


We can easily extend our results to other non-trivial backgrounds,
say, non-compact Calabi-Yau 3-folds, 
just by replacing the $\cN=2$ minimal sector with other
solvable superconformal models. 
The dual gauge theory or the LST interpretations of our results 
and their extension to various setups may be also 
interesting subjects which deserves further studies.
It will be also important to extend our results to
the rolling D-branes with non-vanishing angular momentum in $S^3$, 
which are also discussed in \cite{Kutasov}.
Recalling we can get the trajectory only by rescaling the time
direction, we may construct the boundary states for these D-branes only by
adding a factor in front of the energy $\omega$ in the wave functions
with some proper combinations of the Ishibashi states of $SU(2)$.

To conclude the whole paper, let us comment on 
the possible fate of the rolling D-brane. 
As we can see from the boundary states (or classical analysis), 
the rolling brane inevitably approaches the strong coupling region 
in the far past and future unlike the hairpin brane. 
This has a physical interpretation: It loses all the energy 
and eventually gets absorbed into the NS5-branes, 
making a bound state as was discussed in \cite{Kutasov}. 
To capture this nonperturbative physics is beyond the scope of 
boundary state analysis.
One possible approach to this problem may be to lift the whole setup 
to the M-theory (see \cite{ABKS} in the context of LST). 
Indeed, taking into consideration the success of M-theoretic 
approach to the nonperturbative physics involving 
the static NS5-brane and D-branes ({\em e.g.} Hanany-Witten setup), 
we expect this will be an interesting future direction worth pursuing.


~


\section*{Acknowledgements}
\indent
We would like to thank T. Eguchi, D. Ghoshal,
Y. Imamura, T. Kawano and T. Takayanagi for valuable discussions.

The research of Y. N is supported in part by a Grant
for 21st Century COE Program ``QUESTS'' from the Ministry of Education,
Culture, Sports, Science, and Technology of Japan.
The research of Y. S is also supported by 
the Ministry of Education, Culture, Sports,
Science and Technology of Japan. 
The research of H. T is supported in part by JSPS
Research Fellowships for Young Scientists.

\newpage
\section*{Appendix A ~ Note on $\cN=2$ Minimal Characters}
\setcounter{equation}{0}
\def\theequation{A.\arabic{equation}}

~

A simple way to realize the level $k$ $\cN=2$ minimal model 
$(\hat{c}=k/(k+2))$ is to use the Kazama-Suzuki coset
$\frac{SU(2)_k\times U(1)_2}{U(1)_{k+2}}$. We then have 
the following branching relation;
\begin{eqnarray}
&& \chi_{\ell}^{(k)}(\tau,w)\Th{s}{2}(\tau,w-z)
=\sum_{\stackrel{m\in \bsz_{2(k+2)}}{\ell+m+s\in 2\bsz}} \chi_m^{\ell,s}
(\tau,z)\Th{m}{k+2}(\tau,w-2z/(k+2))~, \nn
&& \chi^{\ell,s}_m(\tau,z) \equiv  0~, ~~~ \mbox{for $\ell+m+s \in 2\bz+1$}~,
\label{branching minimal}
\end{eqnarray}
where $\chi_{\ell}^{(k)}(\tau,z)$ is the spin $\ell/2$ character of 
$SU(2)_k$;
\begin{eqnarray}
&&\chi^{(k)}_{\ell}(\tau, z) 
=\frac{\Th{\ell+1}{k+2}(\tau,z)-\Th{-\ell-1}{k+2}(\tau,z)}
                        {\Th{1}{2}(\tau,z)-\Th{-1}{2}(\tau,z)}
\equiv \sum_{m \in \bsz_{2k}}\, c^{(k)}_{\ell,m}(\tau)\Th{m}{k}(\tau,z)~.
\label{SU(2) character}
\end{eqnarray}
The branching function $\chi^{\ell,s}_m(\tau,z)$ 
is explicitly calculated as follows 
;
\begin{equation}
\chi_m^{\ell,s}(\tau,z)=\sum_{r\in \bsz_k}c^{(k)}_{\ell, m-s+4r}(\tau)
\Th{2m+(k+2)(-s+4r)}{2k(k+2)}(\tau,z/(k+2))~.
\end{equation}
Then, the character formulae of unitary representations 
are written as 
\begin{eqnarray}
&& \ch{(\sNS)}{\ell,m}(\tau,z) = \chi^{\ell,0}_m(\tau,z)
+\chi^{\ell,2}_m(\tau,z)~, \nn
&& \ch{(\stNS)}{\ell,m}(\tau,z) = \chi^{\ell,0}_m(\tau,z)
-\chi^{\ell,2}_m(\tau,z)\equiv 
e^{-i\pi\frac{m}{k+2}}\ch{(\sNS)}{\ell,m}\left(\tau,z+\frac{1}{2}\right)~, \nn
&& \ch{(\sR)}{\ell,m}(\tau,z) = \chi^{\ell,1}_m(\tau,z)
+\chi^{\ell,3}_m(\tau,z) \equiv 
q^{\frac{k}{8(k+2)}}y^{\frac{k}{2(k+2)}}
\ch{(\sNS)}{\ell,m+1}\left(\tau,z+\frac{\tau}{2}\right)~, \nn
&& \ch{(\stR)}{\ell,m}(\tau,z) = \chi^{\ell,1}_m(\tau,z)
-\chi^{\ell,3}_m(\tau,z) \equiv
- e^{-i\pi\frac{m+1}{k+2}}q^{\frac{k}{8(k+2)}}y^{\frac{k}{2(k+2)}}
\ch{(\sNS)}{\ell,m+1}\left(\tau,z+\frac{1}{2}+\frac{\tau}{2}\right)~. \nn
&& \label{minimal character}
\end{eqnarray}
By definition, we may restrict to $\ell+m \in 2\bz$ for the  $\NS$ and
$\tNS$ sectors, and to $\ell+m \in 2\bz+1$ for the $\R$ and $\tR$
sectors.  It is convenient to define 
$\ch{(\sigma)}{*}(\tau,z)\equiv 0$ unless these conditions for $\ell$,
$m$ are satisfied.
Note the character identity (``field identification'')
\begin{eqnarray}
&& \chi^{k-\ell,s+2}_{m+k+2}(\tau,z) = \chi^{\ell,s}_m(\tau,z)~,
\end{eqnarray}
or equivalently, 
\begin{eqnarray}
&& \ch{(\sigma)}{k-\ell,m+k+2}(\tau,z)= \ch{(\sigma)}{\ell,m}(\tau,z)
~,~~(\sigma = \NS,\, \R)~,\nn
&& \ch{(\sigma)}{k-\ell,m+k+2}(\tau,z)= -\ch{(\sigma)}{\ell,m}(\tau,z)
~,~~(\sigma = \tNS,\, \tR)~. 
\label{field identification}
\end{eqnarray}


\end{document}